\documentclass[12pt]{article}%
\usepackage{amsmath}
\usepackage{amsfonts}
\usepackage{amssymb}
\usepackage{graphicx}
\usepackage{hyperref}%
\providecommand{\U}[1]{\protect\rule{.1in}{.1in}}

\begin{document}
\begin{titlepage}
\vspace{.3cm} \vspace{1cm}
\begin{center}
\baselineskip=16pt \centerline{\Large\bf  Cosmology with Mimetic Matter } \vspace{2truecm} \centerline{\large\bf Ali H.
Chamseddine$^{1,2}$, \ Viatcheslav Mukhanov$^{3,4}$, \ Alexander Vikman$^{3}$\ \ } \vspace{.5truecm}
\emph{\centerline{$^{1}$Physics Department, American University of Beirut, Lebanon}}
\emph{\centerline{$^{2}$I.H.E.S. F-91440 Bures-sur-Yvette, France}}
\emph{\centerline{$^{3}$Theoretical Physics, Ludwig Maxmillian University, Theresienstr. 37, 80333 Munich, Germany }}
\emph{\centerline{$^{4}$MPI for Physics, Foehringer Ring, 6, 80850, Munich, Germany}}
\end{center}
\vspace{2cm}
\begin{center}
{\bf Abstract}
\end{center}
We consider minimal extensions of the recently proposed Mimetic Dark Matter and show that by introducing
a potential for the mimetic non-dynamical scalar field we can mimic nearly any gravitational properties of the normal matter. In particular,
the mimetic matter can provide us with inflaton, quintessence and even can lead to a bouncing nonsingular universe.
We also investigate the behaviour of cosmological perturbations due to a mimetic matter.
We demonstrate that simple mimetic inflation can produce red-tilted scalar perturbations which are largely enhanced over gravity waves.
\end{titlepage}

\section{\bigskip Introduction}

Recent work \cite{CM} proposed a modification of general relativity where the
metric $g_{\mu\nu}$ is defined in terms of an auxiliary metric $\tilde{g}%
_{\mu\nu}$ and a scalar field $\phi$ as
\begin{equation}
g_{\mu\nu}=\tilde{g}_{\mu\nu}\ \tilde{g}^{\alpha\beta}\partial_{\alpha}%
\phi\partial_{\beta}\phi\ . \label{1b}%
\end{equation}
The conformal mode of gravity is thus encoded into the scalar field $\phi$.
The theory is manifestly invariant with respect to the Weyl transformations of
the auxiliary metric. The resulting equations of motion are equivalent to the
Einstein equations except for the appearance of an extra longitudinal mode of
the gravitational field which is dynamical even in the absence of normal
matter. It was shown that this extra mode of the gravitational field can serve
as a source of cold Dark Matter. In this paper we generalize this model in a
minimal way and show that Mimetic Matter can imitate practically any
gravitational properties of the normal matter. In particular, this matter can
provide us with quintessence, inflation and even bouncing universe.

The equations of motion derived from the action formulated in terms of the
auxiliary metric $\widetilde{g}_{\mu\nu}$ are equivalent to those ones
obtained by variation of the Einstein action with respect to the physical
metric $g_{\mu\nu}$ if we impose an extra constraint on the scalar field
\cite{Golovnev}, \cite{Barvinsky}. In fact, the theory with action%
\begin{equation}
S=\int d^{4}x\,\sqrt{-g\left(  \tilde{g}_{\mu\nu},\phi\right)  }\left[
-\frac{1}{2}R\left(  g_{\mu\nu}\left(  \tilde{g}_{\mu\nu},\phi\right)
\ \right)  +\mathcal{L}_{m}\right]  ,\label{3n}%
\end{equation}
yields traceless Einstein equations of motion
\begin{equation}
\left(  G^{\mu\nu}-T^{\mu\nu}\ \right)  -\left(  G-T\ \right)  g^{\mu\alpha
}g^{\nu\beta}\partial_{\alpha}\phi\partial_{\beta}\phi=0\ ,\label{einstein}%
\end{equation}
because
\begin{equation}
g^{\mu\nu}\partial_{\mu}\phi\partial_{\nu}\phi=1\ ,\label{3aa}%
\end{equation}
as it follows from definition $\left(  \ref{1b}\right)  .$ This suggests to
use identify (\ref{3aa}) as a constraint by employing a Lagrange multiplier%
\begin{equation}
S=\int d^{4}x\sqrt{-g}\left[  -\frac{1}{2}R\left(  g_{\mu\nu}\right)
+\mathcal{L}_{m}\left(  g_{\mu\nu},...\right)  +\lambda\left(  g^{\mu\nu
}\partial_{\mu}\phi\partial_{\nu}\phi-1\right)  +\tilde{\lambda}\left(
\nabla_{\mu}V^{\mu}-1\right)  \right]  ,\label{1aa}%
\end{equation}
so that both Dark Matter and Dark Energy arise as constants of integration in
the resulting equations. The first constraint here is responsible for the
appearance of mimetic Dark Matter \cite{CM}, while the second one gives the
cosmological constant \cite{HT}. This can be easily seen by examining the
equations of motion of the action (\ref{1aa}) which in addition to imposing
the constraints (\ref{3aa}) and $\nabla_{\mu}V^{\mu}=1$ also give
\begin{align}
G_{\mu\nu}-T_{\mu\nu}+2\lambda\partial_{\mu}\phi\partial_{\nu}\phi+g_{\mu\nu
}\tilde{\lambda} &  =0\ ,\label{Einstein_with_multipl}\\
\partial_{\mu}\tilde{\lambda} &  =0\ .
\end{align}
Thus the cosmological constant $\tilde{\lambda}=\Lambda$ arises as a a
constant of integration, and the Lagrange multiplier $\lambda$ is then
determined from the trace of the Einstein equations
\eqref{Einstein_with_multipl}%
\begin{equation}
\lambda=-\frac{1}{2}(G-T+4\Lambda)\ .
\end{equation}
The metric $g_{\mu\nu}$ and the scalar field $\phi$ are then determined by
equations%
\begin{equation}
\left(  G_{\mu\nu}-T_{\mu\nu}\right)  -\left(  G-T\right)  \partial_{\mu}%
\phi\partial_{\nu}\phi+\left(  g_{\mu\nu}-4\partial_{\mu}\phi\partial_{\nu
}\phi\right)  \Lambda=0\ ,
\end{equation}
together with $\left(  \ref{3aa}\right)  .$ Therefore both Dark Matter and
Dark Energy can arise from the minimal modification of General Relativity (GR)
by adding non-dynamical scalar and vector fields.

The purpose of the present work is to generalize the model above by
introducing an arbitrary potential $V\left(  \phi\right)  $ and study the
cosmological solutions in this theory. In particular, we shall show how the
appropriate choice for the potential $V\left(  \phi\right)  $ can lead to
various cosmological solutions for bouncing universe, inflation and quintessence.

\section{\bigskip Potential for Mimetic Matter}

Consider the theory with the action (compare to \cite{Dust})%
\begin{equation}
S=%
{\displaystyle\int}
d^{4}x\sqrt{-g}\left[  -\frac{1}{2}R\left(  g_{\mu\nu}\right)  +\lambda\left(
g^{\mu\nu}\partial_{\mu}\phi\partial_{\nu}\phi-1\right)  -V\left(
\phi\right)  +\mathcal{L}_{m}\left(  g_{\mu\nu},...\right)  \right]  \ ,
\label{2aa}%
\end{equation}
where we have skipped the cosmological constant constraint because its only
role for the classical solutions is to shift the potential $V\left(
\phi\right)  $ by a constant $\Lambda$. Variation with respect to $\lambda$
gives equation (\ref{3aa}) while varying with respect to $g^{\mu\nu}$ we
obtain
\begin{equation}
G_{\mu\nu}-2\lambda\partial_{\mu}\phi\partial_{\nu}\phi-g_{\mu\nu}V\left(
\phi\right)  =T_{\mu\nu}\ , \label{4}%
\end{equation}
where $G_{\mu\nu}$ and $T_{\mu\nu}$ are the Einstein tensor and the
energy-momentum tensor for the normal matter. Taking trace of these equations
we can express the Lagrange multiplier as
\begin{equation}
\lambda=\frac{1}{2}\left(  G-T-4V\right)  , \label{5}%
\end{equation}
and hence equations $\left(  \ref{4}\right)  $ become%
\begin{equation}
G_{\mu\nu}=\left(  G-T-4V\right)  \partial_{\mu}\phi\partial_{\nu}\phi
+g_{\mu\nu}V\left(  \phi\right)  +T_{\mu\nu} \ , \label{6}%
\end{equation}
which taken together with
\begin{equation}
g^{\mu\nu}\partial_{\mu}\phi\partial_{\nu}\phi=1\ , \label{1a}%
\end{equation}
replace the Einstein equations. They are equivalent to the Einstein equations
with an extra longitudinal degree of freedom. In distinction from General
Relativity, if we take the trace of $\left(  \ref{6}\right)  $ the resulting
equation will be satisfied identically. The missing equation in our case is
replaced by the constraint $\left(  \ref{1a}\right)  $.

We would like to stress that the extra longitudinal degree of freedom of the
gravitational field cannot be entirely attributed to the scalar field $\phi$.
In fact this field satisfies the first order Hamilton-Jacobi type differential
equation and therefore it is not dynamical by itself.

Taking the covariant derivative $\nabla^{\nu}$ of equation (\ref{6}) and using
the Bianchi identity $\nabla^{\nu}G_{\mu\nu}=0$ together with conservation of
the energy-momentum tensor $\nabla^{\nu}T_{\mu\nu}=0,$ we obtain
\begin{equation}
\nabla^{\nu}\left[  \left(  G-T-4V\right)  \partial_{\mu}\phi\partial_{\nu
}\phi+g_{\mu\nu}V\left(  \phi\right)  \right]  \ =0 \ .\ \label{7}%
\end{equation}
This equation can be simplified further if we take into account that%
\begin{equation}
\nabla^{\rho}\left(  g^{\mu\nu}\partial_{\mu}\phi\partial_{\nu}\phi\right)
=2g^{\mu\nu}\left(  \nabla^{\rho}\partial_{\mu}\phi\right)  \partial_{\nu}%
\phi=2g^{\mu\nu}\left(  \nabla_{\mu}\partial^{\rho}\phi\right)  \partial_{\nu
}\phi=0 \ , \label{8}%
\end{equation}
and $\partial_{\mu}\phi\neq0$ at least for one index $\mu$. The result is
\begin{equation}
\nabla^{\nu}\left(  \left(  G-T-4V\right)  \partial_{\nu}\phi\right)
=-V^{\prime}\left(  \phi\right)  , \label{9}%
\end{equation}
where $V^{\prime}\left(  \phi\right)  =\partial V/\partial\phi.$ Equation
$\left(  \ref{9}\right)  $ also follows directly from \ the action $\left(
\ref{2aa}\right)  $ when we vary it with respect to $\phi$ and take into
account $\left(  \ref{5}\right)  .$

It is easy to see that equations $\left(  \ref{6}\right)  $ are equivalent to
the Einstein equations with an extra ideal fluid with pressure%
\begin{equation}
\tilde{p}=-V, \label{9a}%
\end{equation}
and energy density%
\begin{equation}
\tilde{\varepsilon}=G-T-3V. \label{10a}%
\end{equation}
The scalar field $\phi$ plays the role of the velocity potential and the
constraint $\left(  \ref{1a}\right)  $ is simply the normalization condition
for the 4-velocities.

Let us now investigate the solutions of equations $\left(  \ref{6}\right)
,\left(  \ref{1a}\right)  $ in a flat universe with the metric%
\begin{equation}
ds^{2}=dt^{2}-a^{2}\left(  t\right)  \delta_{ik}dx^{i}dx^{k}, \label{10}%
\end{equation}
assuming that ordinary matter is absent, that is, $T_{\mu\nu}=0.$ A general
solution of $\left(  \ref{1a}\right)  $ is in this case%
\begin{equation}
\phi=\pm t+A \ , \label{11a}%
\end{equation}
where $A$ is a constant of integration. Without lose of generality we shall
identify the field $\phi$ with time
\begin{equation}
\phi=t \ .
\end{equation}
Taking into account that the pressure and the energy density, defined in
$\left(  \ref{9a}\right)  $ and $\left(  \ref{10a}\right)  ,$ depend only on
time, equation $\left(  \ref{9}\right)  $ becomes%
\begin{equation}
\frac{1}{a^{3}}\frac{d}{dt}\left(  a^{3}\left(  \tilde{\varepsilon}-V\right)
\right)  =-\dot{V}, \label{12}%
\end{equation}
where we took into account that $V^{\prime}=\dot{V}$, where dot denotes the
derivative with respect to time $t.$ This equation can be easily integrated
and we obtain the following expression for the energy density in terms of the
potential $V,$%
\begin{equation}
\tilde{\varepsilon}=V-\frac{1}{a^{3}}\int a^{3}\dot{V}dt=\frac{3}{a^{3}}\int
a^{2}Vda \ , \label{13}%
\end{equation}
while%
\begin{equation}
\tilde{p}=-V. \label{14}%
\end{equation}
One can easily check that the energy density and the pressure given by these
expressions satisfy the conservation law%
\begin{equation}
\dot{\tilde{\varepsilon}}=-3H\left(  \tilde{\varepsilon}+\tilde{p}\right)  ,
\label{15}%
\end{equation}
where
\[
H\equiv\frac{\dot{a}}{a} \ ,
\]
is the Hubble constant. A constant of integration in $\left(  \ref{13}\right)
$ determines the amount of mimetic dark matter, which decays as $a^{-3}.$ On
the other hand for a nonvanishing $V$ there is an extra contribution to
mimetic matter in the amount entirely fixed by the potential $V$. In this
sense an extra mimetic matter is similar to a cosmological constant added to
the Lagrangian with a fixed value. Therefore, the number of degrees of freedom
in the system does not increase compared to the case of mimetic dust. Because
the field $\phi$ is not a dynamical field, and in Friedmann universe it is a
linear function of time for any $V,$ nothing can prevent us to consider
negative potentials along with positive ones.

The Friedmann equation, obtained from the time-time component of equation
(\ref{6}), takes the form
\begin{equation}
H^{2}=\frac{1}{3}\tilde{\varepsilon}=\frac{1}{a^{3}}\int a^{2}Vda \ ,
\label{16}%
\end{equation}
and for a given $V\left(  \phi\right)  =V\left(  t\right)  $ could be solved
for $a\left(  t\right)  .$ However, instead of solving this integral equation,
it is more convenient to differentiate it first and reduce it to an ordinary
differential equation. Multiplying equation (\ref{16}) by $a^{3}$ and
differentiating it with respect to time we obtain
\begin{equation}
2\dot{H}+3H^{2}=V\left(  t\right)  . \label{17}%
\end{equation}
Note that equation (\ref{17}) could be also obtained from the space-space
component of equation (\ref{6}). It can be simplified further if instead of
$a$ we introduce the new variable
\begin{equation}
y=a^{\frac{3}{2}},
\end{equation}
then
\begin{equation}
H=\frac{2}{3}\frac{\dot{y}}{y},\text{ }\dot{H}=\frac{2}{3}\left(  \frac
{\ddot{y}}{y}-\left(  \frac{\dot{y}}{y}\right)  ^{2}\right)  , \label{18}%
\end{equation}
and equation $\left(  \ref{17}\right)  $ becomes a linear differential
equation
\begin{equation}
\ddot{y}-\frac{3}{4}V\left(  t\right)  y=0\ . \label{19}%
\end{equation}
A\ point which should not be underestimated is that the identification of the
scalar field $\phi$ with time greatly simplifies the problem because for a
given potential the pressure becomes known function of time and $y=a^{3/2}$
satisfies linear differential equation. This allows us easily to find
cosmological solutions.

\section{\bigskip Cosmological solutions}

First we consider the potential%
\begin{equation}
V\left(  \phi\right)  =\frac{\alpha}{\phi^{2}}=\frac{\alpha}{t^{2}}\ ,
\label{20}%
\end{equation}
where $\alpha$ is a constant. The general solution of the equation
\begin{equation}
\ddot{y}-\frac{3\alpha}{4t^{2}}y=0 \ , \label{21}%
\end{equation}
is
\begin{equation}
y=\left\{
\begin{array}
[c]{c}%
C_{1}t^{\frac{1}{2}}\cos\left(  \frac{1}{2}\sqrt{\left\vert 1+3\alpha
\right\vert }\ln t+C_{2}\right)  ,\text{ \ for }\alpha<-1/3 \ ,\\
C_{1}t^{\frac{1}{2}\left(  1+\sqrt{1+3\alpha}\right)  }+C_{2}t^{\frac{1}%
{2}\left(  1-\sqrt{1+3\alpha}\right)  }\ ,\text{\ \ for }\alpha\geq-1/3\ ,
\end{array}
\text{\ \ \ \ \ \ \ \ \ }\right.  \label{22}%
\end{equation}
where $C_{1}$ and $C_{2}$ are constants of integration. It is interesting to
note that for large negative $\alpha$ the solution describes an oscillating
flat universe with singularities and the amplitude of oscillation grows with
time. This, however, is not surprising because this case corresponds to a
large positive pressure. In flat universe the scale factor is defined up to an
overall normalization factor and therefore, assuming $C_{1}\neq0,$ we can
write the general solution for the scale factor in case $\alpha\geq-1/3$ as
\begin{equation}
a\left(  t\right)  =t^{\frac{1}{3}\left(  1+\sqrt{1+3\alpha}\right)  }\left(
1+At^{-\sqrt{1+3\alpha}}\right)  ^{2/3}, \label{23}%
\end{equation}
where $A=C_{2}/C_{1}$ is a constant of integration.

One can substitute this solution in $\left(  \ref{16}\right)  $ to find the
energy density%
\begin{equation}
\tilde{\varepsilon}=3H^{2}=\frac{1}{3t^{2}}\left(  1+\sqrt{1+3\alpha}%
\frac{1-At^{-\sqrt{1+3\alpha}}}{1+At^{-\sqrt{1+3\alpha}}}\right)  ^{2}.
\label{24}%
\end{equation}
The appearance of the constant of integration $A$ is related to the freedom of
having an extra constant of integration in $\left(  \ref{16}\right)  .$ Taking
into account that
\begin{equation}
\tilde{p}=-\frac{\alpha}{t^{2}}\ , \label{25}%
\end{equation}
we will find the equation of state for the mimetic matter%
\begin{equation}
w=\frac{\tilde{p}}{\tilde{\varepsilon}}=-3\alpha\left(  1+\sqrt{1+3\alpha
}\frac{1-At^{-\sqrt{1+3\alpha}}}{1+At^{-\sqrt{1+3\alpha}}}\right)  ^{-2}\ ,
\label{26}%
\end{equation}
In general this equation of state depends on time but in the limit of small
and large $t$ approaches a constant. For $\alpha=-1/3$ we have ultra-hard
equation of state $\tilde{p}=\tilde{\varepsilon}$ and $a\propto t^{1/3}.$ The
case $\alpha=-1/4$ corresponds to ultra-relativistic fluid with $\tilde
{p}=\frac{1}{3}\tilde{\varepsilon}$ at large time and $\tilde{p}%
=3\tilde{\varepsilon}$ when $t\rightarrow0$ if $A\neq0.$ When $\alpha$ is very
small then we have mimetic dark matter with negligible pressure, Finally for
positive $\alpha$ the pressure is negative and if $\alpha\gg1$ the equation of
state approaches the cosmological constant, $\tilde{p}=-\tilde{\varepsilon}.$

In case of an arbitrary power law potential
\begin{equation}
V\left(  \phi\right)  =\alpha\phi^{n}=\alpha t^{n}, \label{27}%
\end{equation}
$n\neq-2,$ the general solution of the differential equation%
\[
\ddot{y}-\frac{3\alpha}{4}t^{n}y=0\ ,
\]
is given in terms of Bessel functions%
\begin{equation}
y=t^{\frac{1}{2}}Z_{\frac{1}{n+2}}\left(  \frac{\sqrt{-3\alpha}}{n+2}%
t^{\frac{n+2}{2}}\right)  . \label{28}%
\end{equation}
If $n<-2,$ that is, the potential decays faster than $1/\phi^{2},$ the
asymptotic at large $t$ is $y\propto t$ and, correspondingly, the scale factor
in the leading order behaves as in dust dominated universe, $a\propto
t^{2/3}.$ For $n>-2$%
\begin{equation}
y\propto t^{-n/4}\exp\left(  \pm i\frac{\sqrt{-3\alpha}}{n+2}t^{\frac{n+2}{2}%
}\right)  , \label{29}%
\end{equation}
as $t\rightarrow\infty.$ Here the behavior of the scale factor drastically
depends on the sign of $\alpha.$ For negative $\alpha,$ corresponding to
positive pressure, the mimetic matter leads to an oscillating universe with
singularities. The case of positive $\alpha$ or negative pressure corresponds
to accelerated, inflationary universe. In particular, for $n=0$ we have an
exponential expansion corresponding to the cosmological constant, while $n=2$
leads to inflationary expansion with%
\begin{equation}
a\propto t^{-1/3}\exp\left(  \sqrt{\frac{\alpha}{12}}t^{2}\right)  ,
\label{30}%
\end{equation}
similar to chaotic inflation with quadratic potential.

\section{Mimetic matter as quintessence}

Let us consider the behavior of mimetic matter in the case when the universe
is dominated by some other matter with constant equation of state
$p=w\varepsilon$ and where the potential is given by
\begin{equation}
V\left(  \phi\right)  =\frac{\alpha}{\phi^{2}}=\frac{\alpha}{t^{2}} \ ,
\end{equation}
In this case the scale factor is
\begin{equation}
a\propto t^{\frac{2}{3\left(  1+w\right)  }} \ , \label{31}%
\end{equation}
and if $\phi=t$ then the energy density of mimetic matter given by $\left(
\ref{13}\right)  $ decays as
\begin{equation}
\tilde{\varepsilon}=-\frac{\alpha}{wt^{2}}\ , \label{32}%
\end{equation}
if we set the constant of integration in $\left(  \ref{13}\right)  $ to zero.
Because $\tilde{p}=-\alpha/t^{2}$ the mimetic matter imitates the equation of
state of the dominant matter. However, since the total energy density is equal
to
\begin{equation}
\varepsilon=3H^{2}=\frac{4}{3\left(  1+w\right)  ^{2}t^{2}}\ , \label{33}%
\end{equation}
this mimetic matter can be subdominant only if $\alpha/w\ll1.$ The more
general solution for subdominant mimetic matter, $\phi=t+t_{0},$ first
corresponds to a cosmological constant for $t<t_{0}$ and only at $t>t_{0\text{
\ }}$ starts to behave similar to a dominant matter.

\section{Mimetic matter as an inflaton\textit{ }}

One can easily construct the inflationary solutions using the mimetic matter.
In fact, one can take any scale factor $a\left(  t\right)  =y^{2/3}$ and using
$\left(  \ref{21}\right)  $ find the potential%
\begin{equation}
V\left(  \phi\right)  =V\left(  t\right)  =\frac{4}{3}\frac{\ddot{y}}{y} \ ,
\label{34}%
\end{equation}
for the theory where this scale factor will be a solution of the corresponding
equations. For example, the potential%
\begin{equation}
V\left(  \phi\right)  =\frac{\alpha\phi^{2}}{\exp\left(  \phi\right)  +1}\ ,
\label{35}%
\end{equation}
with positive $\alpha$ describes inflation with graceful exit to matter
dominating universe. In fact, the scale factor grows as%
\begin{equation}
a\propto\exp\left(  -\sqrt{\frac{\alpha}{12}}t^{2}\right)  \ , \label{36}%
\end{equation}
at large negative $\phi=t$ and it is proportional to $t^{2/3}$ for positive
$t.$ Playing with potentials one can easily get any \textquotedblleft
wishful\textquotedblright\ behavior for the scale factor during inflation and
after it. Thus we see that the mimetic matter can easily provide us with an
inflaton. The question is then how one can generate the radiation and baryons
we observe. This can be done either via gravitational particle production at
the end of inflation \cite{Ford, Damour, Peebles}, or via direct coupling of
the other fields to $\phi.$ We will leave it to the reader to elaborate this
question, but there are no obvious obstacles which would prevent us from
making the models of inflation using the mimetic matter.

\section{Mimetic matter and bouncing universe}

Using the same strategy as for building inflationary solutions we can easily
find a theory which provide us with non-singular bounce in a contracting flat
universe. For example, let us consider the potential%
\begin{equation}
V\left(  \phi\right)  =\frac{4}{3}\frac{1}{\left(  1+\phi^{2}\right)  ^{2}%
}=\frac{4}{3}\frac{1}{\left(  1+t^{2}\right)  ^{2}} \ . \label{37}%
\end{equation}
The general exact solution of the equation
\begin{equation}
\frac{d^{2}y}{dt^{2}}-\frac{1}{\left(  t^{2}+1\right)  ^{2}}y=0 \ , \label{38}%
\end{equation}
is
\begin{equation}
y=\sqrt{t^{2}+1}\left(  C_{1}+C_{2}\arctan t\right)  , \label{39}%
\end{equation}
and correspondingly the scale factor is%
\begin{equation}
a=\left(  \sqrt{t^{2}+1}\left(  1+A\arctan t\right)  \right)  ^{2/3}\ ,
\label{40}%
\end{equation}
where we have assumed that $C_{1}\neq0$ and used the freedom for normalization
of the scale factor in a flat universe. Let us consider for simplicity $A=0.$
In this case
\begin{equation}
a=\left(  t^{2}+1\right)  ^{1/3}\ . \label{41}%
\end{equation}
The energy density and pressure are%
\begin{equation}
\tilde{\varepsilon}=3H^{2}=\frac{4}{3}\frac{t^{2}}{\left(  1+t^{2}\right)
^{2}},\text{ \ \ }\tilde{p}=-\frac{4}{3}\frac{1}{\left(  1+t^{2}\right)  ^{2}%
}, \label{42}%
\end{equation}
respectively. For large negative $t$ the universe is dominated by dust with
negligible pressure and it contracts. The energy density first grows as
$a^{-3}$ and then during a very short time interval around $\left\vert
t\right\vert $ $\sim1$ it drastically drops to zero, the universe stops
contraction and begins to expand. After the beginning of expansion the energy
density first increases to the Planckian value within short interval
corresponding to the Planckian time and then the expansion proceeds as in dust
dominated universe. The interesting property of the model is the change of
equation of state from the normal one, satisfying condition $\tilde
{\varepsilon}+\tilde{p}>0$ to the phantom one $\tilde{\varepsilon}+\tilde
{p}<0.$ In fact,%
\begin{equation}
\tilde{\varepsilon}+\tilde{p}=\frac{4}{3}\frac{t^{2}-1}{\left(  1+t^{2}%
\right)  ^{2}}\ , \label{43}%
\end{equation}
and for $\left\vert t\right\vert <1$ the mimetic matter is a phantom and
therefore we can have a non-singular bounce. In the general case for $A\neq0,$
the bounce is non-singular if $\left\vert A\right\vert <2/\pi,$ otherwise the
contracting universe ends up in the singularity. The value of $A$ is
determined by the relative balance of phantom and dark matter
\textquotedblleft components\textquotedblright\ in mimetic matter when we are
close to the point of bounce. Therefore it is not surprising that the universe
becomes singular when dark matter dominates. Using this phantom mimetic matter
one can also avoid a singularity in contracting universe in the presence of
other matter.

In the model we have considered the bounce happens at Planckian scales when
the theory is not under control because of quantum gravitational effects.
However slightly modifying the potential $V$ we can easily lower the bounce
scale and make the bounce duration to be longer than the Planckian time. In
fact let us consider the potential
\begin{equation}
V\left(  \phi\right)  =\frac{4}{3}\frac{\alpha}{\left(  \phi_{0}^{2}+\phi
^{2}\right)  ^{2}}=\frac{4}{3}\frac{\alpha}{\left(  t_{0}^{2}+t^{2}\right)
^{2}}\ . \label{44}%
\end{equation}
For this potential \ the differential equation $\left(  \ref{19}\right)  $
becomes%
\begin{equation}
\frac{d^{2}y}{d\tilde{t}^{2}}-\frac{\alpha t_{0}^{-2}}{\left(  \tilde{t}%
^{2}+1\right)  ^{2}}y=0\ , \label{45}%
\end{equation}
where $\tilde{t}=t/t_{0},$ and its general solution is
\begin{equation}
a\left(  t\right)  =y^{\frac{2}{3}}=\left(  \sqrt{\left(  \frac{t}{t_{0}%
}\right)  ^{2}+1}\left(  \cos\left(  \beta\arctan\frac{t}{t_{0}}\right)
+A\sin(\beta\arctan\frac{t}{t_{0}})\right)  \right)  ^{\frac{2}{3}},
\label{46}%
\end{equation}
where $\beta=\sqrt{1-\alpha t_{0}^{-2}}.$ In this case the bounce happens at
scales about $\alpha t_{0}^{-2}$ during the time interval $t_{0}.$

\section{Cosmological perturbations}

Finally we will consider the behavior of small longitudinal metric
perturbations in the universe dominated by mimetic matter. Because the off-
diagonal components of the energy-momemtum tensor for mimetic matter vanish in
the linear order, the metric of perturbed universe in the Newtonian gauge can
be written as \cite{Mukhanov}%
\begin{equation}
ds^{2}=\left(  1+2\Phi\right)  dt^{2}-\left(  1-2\Phi\right)  a^{2}\delta
_{ik}dx^{i}dx^{k}, \label{47}%
\end{equation}
where $\Phi$ is the Newtonian gravitational potential. Considering
perturbations of the scalar field,%
\begin{equation}
\phi=t+\delta\phi\ , \label{48}%
\end{equation}
from equation $\left(  \ref{1a}\right)  $ we find
\begin{equation}
\Phi=\delta\dot{\phi}\ . \label{49}%
\end{equation}
The equation for perturbations which follows from the linearized $0-i$
components of Einstein equations is (see equations (7.39) and (7.45) in
\cite{Mukhanov})%
\begin{equation}
\left(  \dot{\Phi}+H\Phi\right)  _{,i}=\frac{1}{2}\left(  \tilde{\varepsilon
}+\tilde{p}\right)  \delta\phi_{,i}\ . \label{50}%
\end{equation}
Taking into account that $\tilde{\varepsilon}+\tilde{p}=-2\dot{H}$ and
substituting here $\Phi$ from $\left(  \ref{49}\right)  $ we obtain the
following equation for $\delta\phi$%
\begin{equation}
\delta\ddot{\phi}+H\delta\dot{\phi}+\dot{H}\delta\phi=0\ . \label{51}%
\end{equation}
As one can easily verify by direct substitution the general solution of this
equation is%
\begin{equation}
\delta\phi=A\frac{1}{a}\int adt\ , \label{52}%
\end{equation}
where $A$ is a constant of integration which depends only on the spatial
coordinates (the other constant of integration corresponding to decaying mode
can be always included in the integral). The corresponding gravitational
potential is%
\begin{equation}
\Phi=\delta\dot{\phi}=A\frac{d}{dt}\left(  \frac{1}{a}\int adt\right)
=A\left(  1-\frac{H}{a}\int adt\right)  . \label{53}%
\end{equation}
This is exactly the general solution we normally have for the long wavelength
cosmological perturbations when one can neglect the spatial derivative terms
which are multiplied by the speed of sound for normal hydrodynamical fluid
\cite{Mukhanov}. However, in our case the solution above is universally valid
for all perturbations irrespective of their wavelength. In this sense the
perturbations behave as a dust with vanishing speed of sound even for mimetic
matter with nonvanishing pressure (in agreement with \cite{Dust}). As a result
one cannot define the quantum fluctuations of mimetic matter in the usual way.
If so the mimetic inflation considered above would fail in one of its major
tasks, namely, in explaining of the large scale structure as originated from
quantum fluctuations. To make the mimetic inflation \textquotedblleft
viable\textquotedblright\ one either has to use one more scalar field playing
the role of curvaton \cite{Curva} or slightly modify the model for mimetic
matter. The use of extra scalar field makes the theory not very plausible
because such a theory can \textquotedblleft explain\textquotedblright\ nearly
everything and predict nothing. For this reason we prefer to modify the model above.

\section{Modified mimetic matter action and cosmological perturbations}

Let us add to the Lagrangian $\left(  \ref{2aa}\right)  $ an extra term%
\begin{equation}
+\frac{1}{2}\gamma\left(  \square\phi\right)  ^{2}, \label{54}%
\end{equation}
where $\gamma$ is a constant and $\square=g^{\mu\nu}\nabla_{\mu}\nabla_{\nu}$.
Note that because the scalar field satisfies the constraint%
\begin{equation}
g^{\mu\nu}\partial_{\mu}\phi\partial_{\nu}\phi=1\ , \label{55}%
\end{equation}
by adding the terms with many derivatives of $\phi$ we do not change the total
number of degrees of freedom. Varying the action
\begin{equation}
S=%
{\displaystyle\int}
d^{4}x\sqrt{-g}\left[  -\frac{1}{2}R\left(  g_{\mu\nu}\right)  \ +\lambda
\left(  g^{\mu\nu}\partial_{\mu}\phi\partial_{\nu}\phi-1\right)  -V\left(
\phi\right)  +\frac{1}{2}\gamma\left(  \square\phi\right)  ^{2}\right]  ,
\label{56}%
\end{equation}
with respect to the metric we obtain
\begin{equation}
G_{\nu}^{\mu}=\tilde{T}_{\nu}^{\mu}\ , \label{57}%
\end{equation}
where
\begin{equation}
\tilde{T}_{\nu}^{\mu}=\left(  V+\gamma\left(  \phi_{,\alpha}\chi^{,\alpha
}+\frac{1}{2}\chi^{2}\right)  \right)  \delta_{\nu}^{\mu}+2\lambda\phi_{,\nu
}\phi^{,\mu}-\gamma\left(  \phi_{,\nu}\chi^{,\mu}+\chi_{,\nu}\phi^{,\mu
}\right)  \ , \label{58}%
\end{equation}
and
\begin{equation}
\chi=\square\phi\ . \label{59}%
\end{equation}
Taken together with constraint $\left(  \ref{55}\right)  $ the equations
$\left(  \ref{57}\right)  $ completely determine the metric, scalar field and
Lagrange multiplier $\lambda.$ Instead of making analogy with perfect fluid it
is more convenient to solve equations $\left(  \ref{57}\right)  $ directly.
The general solution of equation $\left(  \ref{55}\right)  $ in the Friedmann
universe is
\begin{equation}
\phi=t+A\ , \label{60}%
\end{equation}
and hence
\begin{equation}
\chi=\square\phi=\ddot{\phi}+3H\dot{\phi}=3H\ . \label{61}%
\end{equation}
Taking this into account, the $0-0$ Einstein equation then reduces to%
\begin{equation}
H^{2}=\frac{1}{3}V+\gamma\left(  \frac{3}{2}H^{2}-\dot{H}\right)  +\frac{2}%
{3}\lambda\ , \label{62}%
\end{equation}
and the $i-j$ equations give%
\begin{equation}
2\dot{H}+3H^{2}=V\left(  t\right)  +\frac{3\gamma}{2}\left(  2\dot{H}%
+3H^{2}\right)  \ . \label{63}%
\end{equation}
Thus, instead of equation $\left(  \ref{17}\right)  $ we obtain
\begin{equation}
2\dot{H}+3H^{2}=\frac{2}{2-3\gamma}V, \label{64}%
\end{equation}
which is different from $\left(  \ref{17}\right)  $ only by overall
normalization of potential $V.$ Therefore, in the presence of the extra
$\left(  \square\phi\right)  ^{2}$ term the cosmological solutions derived
above for homogeneous universe will remain unchanged up to the numerical
factor of order unity. However, this term changes the behavior of the short
wave cosmological perturbations drastically. The linear perturbation of $0-i$
component of the energy momentum tensor is equal to
\begin{equation}
\delta T_{i}^{0}=2\lambda\delta\phi_{,i}-3\gamma\dot{H}\delta\phi_{,i}%
-\gamma\delta\chi_{,i}\ . \label{65}%
\end{equation}
Taking into account that
\begin{align}
\delta\chi &  =\delta\left(  \square\phi\right)  =-4\dot{\Phi}-6H\Phi
+\delta\ddot{\phi}+3H\delta\dot{\phi}-\frac{\Delta}{a^{2}}\delta
\phi=\nonumber\\
&  =-3\delta\ddot{\phi}-3H\delta\dot{\phi}-\frac{\Delta}{a^{2}}\delta\phi\ ,
\label{67}%
\end{align}
and
\begin{equation}
\lambda=\left(  3\gamma-1\right)  \dot{H}\ , \label{68}%
\end{equation}
as it follows from equations $\left(  \ref{62}\right)  $ and $\left(
\ref{63}\right)  $, we find that the perturbed $0-i$ Einstein equation reduces
to%
\begin{equation}
\delta\ddot{\phi}+H\delta\dot{\phi}-\frac{c_{s}^{2}}{a^{2}}\Delta\delta
\phi+\dot{H}\,\delta\phi=0\ , \label{69}%
\end{equation}
where
\begin{equation}
c_{s}^{2}=\frac{\gamma}{2-3\gamma}\ . \label{70}%
\end{equation}
This equation is different from $\left(  \ref{51}\right)  $ only by the
presence of the gradient terms multiplied by the speed of sound $c_{s}.$ For a
plane wave perturbation $\propto\exp\left(  ikx\right)  $ equation $\left(
\ref{69}\right)  $ becomes%
\begin{equation}
\delta\phi_{k}^{\prime\prime}+\left(  c_{s}^{2}k^{2}+\frac{a^{\prime\prime}%
}{a}-2\left(  \frac{a^{\prime}}{a}\right)  ^{2}\right)  \delta\phi_{k}=0\ .
\label{71}%
\end{equation}
where prime denotes the derivative with respect to conformal time $\eta=\int
dt/a.$ Considering the short wavelength perturbations with $c_{s}k\eta\gg1$
($\lambda_{ph}=a/k\ll c_{s}H^{-1}$) we can neglect the time derivative terms
inside the bracket and the corresponding solution is%
\begin{equation}
\delta\phi_{k}\propto e^{\pm ic_{s}k\eta}\ .\text{ } \label{72}%
\end{equation}
On the other hand for the long wavelength perturbations with $c_{s}k\eta\ll1$
($\lambda_{ph}=a/k\gg c_{s}H^{-1}$) the $c_{s}^{2}k^{2}-$term can be neglected
and we obtain the same solution as in $\left(  \ref{52}\right)  ,$ that is,
\begin{equation}
\delta\phi=A\frac{1}{a}\int a^{2}d\eta\ . \label{73}%
\end{equation}
To fix the amplitude of quantum fluctuations we have to identify the canonical
quantization variable. Expanding action to the second order in perturbations
gives
\begin{equation}
S=-\frac{1}{2}\int d\eta d^{3}x\left(  \frac{\gamma}{c_{s}^{2}}\delta
\phi^{\prime}\Delta\delta\phi^{\prime}+...\right)  \ , \label{74}%
\end{equation}
and hence for the short wavelength quantum initial perturbations the
canonically normalized variable is
\begin{equation}
v_{k}\sim\frac{\sqrt{\gamma}}{c_{s}}\ k\ \delta\phi_{k}\ , \label{75}%
\end{equation}
whose fluctuation in vacuum is
\begin{equation}
\delta v_{k}\sim\frac{1}{\sqrt{\omega_{k}}}\sim\frac{1}{\sqrt{c_{s}k}}\ ,
\label{VacuumV}%
\end{equation}
so that consequently
\[
\delta\phi_{k}\sim\sqrt{\frac{c_{s}}{\gamma}}\ k^{-3/2}\ .
\]
This implies a flat power-spectrum for $\delta\phi_{\lambda}$ and
$\Phi_{\lambda}\sim\lambda^{-1}$ for short length-scales $\lambda=1/k$. Let us
consider an inflationary stage when%
\begin{equation}
\frac{1}{a}\int adt\simeq H^{-1}. \label{76}%
\end{equation}
Taking this into account and matching the solution \eqref{72} and \eqref{73}
at $c_{s}k\eta\sim1$ we find that%
\begin{equation}
A_{k}\sim\sqrt{\frac{c_{s}}{\gamma}}\ \frac{H_{c_{s}k\sim Ha}}{k^{3/2}}\ .
\end{equation}
Therefore the typical amplitude of the gravitational potential $\left(
\Phi_{k}\sim A_{k}\right)  $ in comoving scales $\lambda\sim1/k$ after
inflation is%
\begin{equation}
\Phi_{\lambda}\sim A_{k}\ k^{3/2}\sim\sqrt{\frac{c_{s}}{\gamma}}%
\ H_{c_{s}k\sim Ha}\ . \label{77}%
\end{equation}
Similarly to \emph{k}-inflation \cite{K_Infl,K_Infl_Pert} one obtains that
\begin{equation}
\text{for}\text{ \ }c_{s}\ll1\ :\ \ \Phi_{\lambda}\sim c_{s}^{-1/2}%
\,H_{c_{s}k\sim Ha}\ , \label{Phi_sublum}%
\end{equation}
which is $c_{s}^{-1/2}$ enhanced with respect to the amplitude of the gravity
waves $h_{\lambda}\sim H_{k\sim Ha}$. On the other hand,
\begin{equation}
\text{for \ }c_{s}\gg1\ :\ \ \Phi_{\lambda}\sim c_{s}^{1/2}\,H_{c_{s}k\sim
Ha}\ , \label{Phi_superlum}%
\end{equation}
which is again $c_{s}^{1/2}$ enhanced\footnote{Note that this superluminality
does not imply any causal paradoxes or other inconsistencies on the level of
effective field theory, see e.g. \cite{Superlum}.} with respect to the
amplitude of the gravity waves. This is a completely opposite effect to what
happens in the \emph{k}-inflationary case where for $c_{s}\gg1$ the scalar
fluctuations are suppressed similarly to \eqref{Phi_sublum} and effectively
the gravity waves are enhanced, see e.g. \cite{LargeGW}. In fact in this
mimetic model the scalar perturbations are always larger than the tensor perturbations.

Here it is important to mention that this suppression of the gravity waves
occurs purely due to the quadratic term with the d'Alembertian. Thus, one can
expect that, contrary to \emph{k}-inflation, this suppression does not induce
any non-Gaussianity.

We would like to stress that similar to usual inflation the spectral index for
the adiabatic fluctuations is also red-tilted for the mimetic inflation.

\section{Conclusions}

In this paper we have extended the Mimetic Dark Matter to mimic any background
cosmology. This can be achieved by adding an appropriate potential $V(\phi)$
to the original metric. As simple examples we have discussed quintessence,
inflation and bouncing universe with vanishing speed of sound for
perturbations.\newline Further, we have found another interesting novel
extension which allows for the nontrivial speed of sound. This can be achieved
by adding higher-order-derivative terms to the action without increasing the
number of degrees of freedom in the system. This allows one to quantize the
inflationary scalar perturbations using standard techniques. It is
demonstrated that these perturbations can have novel observational features
absent in the case of \emph{k}-inflation models. In particular it is possible
to strongly suppress the gravitational waves from inflation, seemingly without
any non-Gaussianity. It would be very interesting to analyze whether one can
observationally distinguish Mimetic Inflation from other models.

Finally, such a modification opens up a new interesting playground for
modeling Dark Matter, where the speed of sound can be very small but not
exactly vanishing and the behavior of the mimetic matter can deviate from the
usual perfect-fluid-like dust.

\subsection*{Acknowledgements}

\bigskip The work of AHC is supported in part by the National Science
Foundation Phys-1202671. The work of VM is supported by TRR 33
\textquotedblleft The Dark Universe\textquotedblright\ and the Cluster of
Excellence EXC 153 \textquotedblleft Origin and Structure of the
Universe\textquotedblright. \bigskip\


\begin{thebibliography}{99}                                                                                               %


\bibitem {CM}A.~H. Chamseddine and V.~Mukhanov, \textit{{Mimetic Dark Matter}%
}, \href{http://dx.doi.org/10.1007/JHEP11(2013)135}{\emph{JHEP}
{\bfseries 1311} (2013) 135},
\href{http://arxiv.org/abs/1308.5410}{{\ttfamily arXiv:1308.5410
[astro-ph.CO]}}.




\bibitem {Golovnev}A.~Golovnev, \textit{{On the recently proposed Mimetic Dark
Matter}}, \href{http://dx.doi.org/10.1016/j.physletb.2013.11.026}{\emph{Phys.
Lett.} {\bfseries B728} (2014) 39--40},
\href{http://arxiv.org/abs/1310.2790}{{\ttfamily arXiv:1310.2790 [gr-qc]}}.




\bibitem {Barvinsky}A.~Barvinsky, \textit{{Dark matter as a ghost free
conformal extension of Einstein theory}},
\href{http://arxiv.org/abs/1311.3111}{{\ttfamily arXiv:1311.3111 [hep-th]}}.




\bibitem {HT}M.~Henneaux and C.~Teitelboim, \textit{{The Cosmological Constant
and General Covariance}},
\href{http://dx.doi.org/10.1016/0370-2693(89)91251-3}{\emph{Phys.Lett.}
{\bfseries B222} (1989) 195--199}.




\bibitem {Dust}E.~A. Lim, I.~Sawicki, and A.~Vikman, \textit{{Dust of Dark
Energy}}, \href{http://dx.doi.org/10.1088/1475-7516/2010/05/012}{\emph{JCAP}
{\bfseries
1005} (2010) 012},
\href{http://arxiv.org/abs/1003.5751}{{\ttfamily arXiv:1003.5751
[astro-ph.CO]}}.




\bibitem {Mukhanov}V.~Mukhanov,
\href{http://www.amazon.com/Physical-Foundations-Cosmology-Viatcheslav-Mukhanov/dp/0521563984}{\emph{{Physical
foundations of cosmology}}}. \newblock Cambridge University Press, 2005.



\bibitem {Ford}L.~Ford, \textit{{Gravitational Particle Creation and
Inflation}},
\href{http://dx.doi.org/10.1103/PhysRevD.35.2955}{\emph{Phys.Rev.}
{\bfseries
D35} (1987) 2955}.




\bibitem {Damour}T.~Damour and A.~Vilenkin, \textit{{String theory and
inflation}},
\href{http://dx.doi.org/10.1103/PhysRevD.53.2981}{\emph{Phys.Rev.}
{\bfseries
D53} (1996) 2981--2989},
\href{http://arxiv.org/abs/hep-th/9503149}{{\ttfamily arXiv:hep-th/9503149
[hep-th]}}.




\bibitem {Peebles}P.~Peebles and A.~Vilenkin, \textit{{Quintessential
inflation}},
\href{http://dx.doi.org/10.1103/PhysRevD.59.063505}{\emph{Phys.Rev.}
{\bfseries D59} (1999) 063505},
\href{http://arxiv.org/abs/astro-ph/9810509}{{\ttfamily arXiv:astro-ph/9810509
[astro-ph]}}.




\bibitem {Superlum}E.~Babichev, V.~Mukhanov, and A.~Vikman,
\textit{{k-Essence, superluminal propagation, causality and emergent
geometry}}, \href{http://dx.doi.org/10.1088/1126-6708/2008/02/101}{\emph{JHEP}
{\bfseries
02} (2008) 101},
\href{http://arxiv.org/abs/0708.0561}{{\ttfamily arXiv:0708.0561 [hep-th]}}.




\bibitem {Curva}A.~D. Linde and V.~F. Mukhanov, \textit{{Nongaussian
isocurvature perturbations from inflation}},
\href{http://dx.doi.org/10.1103/PhysRevD.56.R535}{\emph{Phys.Rev.}
{\bfseries D56} (1997) 535--539},
\href{http://arxiv.org/abs/astro-ph/9610219}{{\ttfamily arXiv:astro-ph/9610219
[astro-ph]}}.




\bibitem {K_Infl}C.~Armendariz-Picon, T.~Damour, and V.~F. Mukhanov,
\textit{{k-inflation}},
\href{http://dx.doi.org/10.1016/S0370-2693(99)00603-6}{\emph{Phys. Lett.}
{\bfseries B458} (1999) 209--218},
\href{http://arxiv.org/abs/hep-th/9904075}{{\ttfamily arXiv:hep-th/9904075}}.




\bibitem {K_Infl_Pert}J.~Garriga and V.~F. Mukhanov, \textit{{Perturbations in
k-inflation}},
\href{http://dx.doi.org/10.1016/S0370-2693(99)00602-4}{\emph{Phys. Lett.}
{\bfseries B458} (1999) 219--225},
\href{http://arxiv.org/abs/hep-th/9904176}{{\ttfamily arXiv:hep-th/9904176}}.




\bibitem {LargeGW}V.~F. Mukhanov and A.~Vikman, \textit{{Enhancing the
tensor-to-scalar ratio in simple inflation}},
\href{http://dx.doi.org/10.1088/1475-7516/2006/02/004}{\emph{JCAP}
{\bfseries
0602} (2006) 004},
\href{http://arxiv.org/abs/astro-ph/0512066}{{\ttfamily arXiv:astro-ph/0512066
[astro-ph]}}.

\end{thebibliography}
\end{document}